\documentclass{aa}

\usepackage{amsmath,amssymb}
\usepackage{siunitx}
\usepackage[colorlinks=true,allcolors=blue]{hyperref}
\usepackage{cleveref}
\usepackage{xspace}
\usepackage{txfonts}
\usepackage[english]{babel}
\usepackage{graphicx, subfig}
\usepackage[normalem]{ulem}
\usepackage{verbatim}
\usepackage{multirow}
\usepackage{mathtools}
\usepackage{epstopdf}
\usepackage{url}
\usepackage{xcolor,color}
\usepackage{orcidlink}
\usepackage{natbib,twoopt}
\usepackage[hyphenbreaks]{breakurl}
\usepackage{booktabs}
\usepackage{placeins}

\crefformat{equation}{eq.~#2#1#3}
\crefformat{figure}{Fig.~#2#1#3}
\crefformat{section}{Sec.~#2#1#3}
\crefformat{subsection}{Sec.~#2#1#3}
\crefformat{table}{Tab.~#2#1#3}

\newcommand{\dd}{\mathop{\mathrm{d}\!}{}}
\newcommand{\deriv}[2]{\dfrac{\dd #1}{\dd #2}}

\newcommand{\ped}[1]{_{\rm #1}}
\newcommand{\lcdm}{$\Lambda$CDM\xspace}

\newcommand{\muv}{{\rm M\ped{UV}}}
\newcommand{\Msun}{{\rm M_{\odot}}}

\def\be{\begin{equation}}
\def\ee{\end{equation}}

\bibpunct{(}{)}{;}{a}{}{,}             
\definecolor{cobalt}{rgb}{0.06, 0.2, 0.65}
\hypersetup{
  colorlinks,
  citecolor=cobalt,
  linkcolor=[rgb]{0.8, 0.2, 1.0},
  urlcolor=cobalt,
}
\makeatletter
  \newcommandtwoopt{\citeads}[3][][]{\href{http://adsabs.harvard.edu/abs/#3}%
    {\def\hyper@linkstart##1##2{}%
     \let\hyper@linkend\@empty\citealp[#1][#2]{#3}}}
  \newcommandtwoopt{\citepads}[3][][]{\href{http://adsabs.harvard.edu/abs/#3}%
    {\def\hyper@linkstart##1##2{}%
     \let\hyper@linkend\@empty\citep[#1][#2]{#3}}}
  \newcommandtwoopt{\citetads}[3][][]{\href{http://adsabs.harvard.edu/abs/#3}%
    {\def\hyper@linkstart##1##2{}%
     \let\hyper@linkend\@empty\citet[#1][#2]{#3}}}
  \newcommandtwoopt{\citeyearads}[3][][]%
    {\href{http://adsabs.harvard.edu/abs/#3}
    {\def\hyper@linkstart##1##2{}%
     \let\hyper@linkend\@empty\citeyear[#1][#2]{#3}}}
\makeatother

\def\apjl{ApJL}
\def\apjs{ApJS}
\def\aap{A\&A}

\def\Msun{\rm M_\odot}

\def\Lsun{{\rm L}_\odot}

\newcommand{\fpbh}{f\ped{PBH}}

\newcommand{\Mpbh}{M\ped{PBH}}
\newcommand{\spbh}{\sigma\ped{PBH}}

\newcommand\code[1]{\textsc{\MakeLowercase{#1}}}
\graphicspath{{plots/}}

\linenumbers
\renewcommand\makeLineNumber{}

\defcitealias{ferrara:2023}{AFM}
\def\AFM{\citetalias{ferrara:2023}\xspace}
\defcitealias{dekel:2023}{\rm FFB}
\def\FFB{\citetalias{dekel:2023}\xspace}
\defcitealias{mason:2023}{BSF}
\def\BSF{\citetalias{mason:2023}\xspace}
\defcitealias{matteri:2025}{PBH}
\def\PBH{\citetalias{matteri:2025}\xspace}

\graphicspath{{plots/}}

\title{Clustering constraints on super-early galaxy formation scenarios}
\titlerunning{Clustering constraints on super-early galaxies}

\author{
Antonio Matteri$^{1}$ \orcidlink{0009-0007-9985-9112} \and
Andrea Pallottini$^{2}$ \orcidlink{0000-0002-7129-5761} \and
Andrea Ferrara$^{1}$ \orcidlink{0000-0002-9400-7312}
       }
\authorrunning{Matteri et al.}
\institute{
Scuola Normale Superiore, Piazza dei Cavalieri 7, 56126 Pisa, Italy
\and Dipartimento di Fisica ``Enrico Fermi'', Universit\`{a} di Pisa, Largo Bruno Pontecorvo 3, Pisa I-56127, Italy
}

\date{Received May 8, 2026; accepted MMMM DDD, YYYY}

\abstract{
The unexpectedly high abundance of bright, blue, super-early galaxies ($z\gtrsim 10$) has challenged most pre-JWST models of early galaxy formation and motivated a wide range of proposed explanations.
We systematically investigate whether galaxy clustering can discriminate among representative scenarios that reproduce the observed UV luminosity function.
Using the Shin-Uchuu dark-matter-only simulation, we populate $z\approx 11$ halos with galaxies according to solutions based on \textit{i)} attenuation-free, \textit{ii)} feedback-free bursts, \textit{iii)} bursty star formation, and \textit{iv)} primordial black hole models.
For each model, we compute the two-point correlation function and predict the galaxy bias for flux-limited samples at different thresholds in the $-20 < \rm M_{\rm UV} < -16$ magnitude range.
We find that all models predict similar bias values ($b\approx 7$) for faint galaxies ($\rm M_{\rm UV} \approx -16$), but diverge at $\rm M_{\rm UV} \lesssim -18$, as the underlying halo-mass to $\rm M_{\rm UV}$ relations differ significantly.
In particular, the primordial black hole scenario predicts an almost luminosity-independent bias, whereas the other models generally predict increasing bias with luminosity, reaching $b\approx 14$ for $\rm M_{\rm UV}\approx -19$. 
Current observational estimates of the bias cannot yet rule out any of the models at a significant statistical confidence.
More precise measurements from future JWST programs, together with improved theoretical predictions, will be required to break the present degeneracies. Ideally, constraints from a complete sample of galaxies with $\rm M_{\rm UV}<-18$ would probe the knee of the $b(\rm M_{\rm UV})$ function, taking advantage of the difference in model predictions and strengthening our analysis.
Although requiring further refinement, galaxy clustering is confirmed to be a promising probe of the physical origin of the JWST high-redshift luminosity function.
}

\keywords{Galaxies: high-redshift -- statistics}

\begin{document}

\maketitle

\section{Introduction}

The James Webb Space Telescope (JWST, first light in 2022) has revolutionized our knowledge of the early Universe.
Its near infrared camera (NIRCam) has been used to photometrically detected several galaxies at $z>10$ \citep{castellano:2022,finkelstein:2022,naidu:2022,naidu:2022a,adams:2023,atek:2023,donnan:2023,donnan:2023a,finkelstein:2023,furtak:2023,harikane:2023,leung:2023,robertson:2023, santini:2023,whitler:2023,yan:2023,casey:2024,hainline:2024,merlin:2024,conselice:2025}, estimate their abundance \citep[][]{donnan:2024, finkelstein:2024, mcleod:2024, robertson:2024, whitler:2025, chemerynska:2026}, and possibly even discover objects at $z>15$ \citep{castellano:2025,perez-gonzalez:2025}.
Its near infrared spectrograph (NIRSpec), instead, has been used to spectroscopically analyze and confirm several high-$z$ galaxies \citep{arrabalharo:2022,arrabalharo:2023,bunker:2023,curtis-lake:2023,fujimoto:2023,tacchella:2023,wang:2023,castellano:2024,hsiao:2024,zavala:2025}, up to $z \approx 14$ \citep{carniani:2024,naidu:2025}.

Thanks to the unprecedented spectral coverage of JWST, these observations probed the statistical and physical properties of galaxies in the first billion years after the Big Bang. The abundance of bright objects, quantified by the ultra-violet (UV) luminosity function (LF), measured at $z \gtrsim 10$ is dramatically higher than most pre-JWST predictions.
Most numerical simulations \citep[e.g.][]{wilkins:2017,kannan:2022}, semi-analytical models \citep[e.g.][]{dayal:2014,mason:2015,yung:2020}, and extrapolations from empirical fits at lower $z$ \citep[e.g.][]{bouwens:2022} predicted an abundance about 10 times smaller than the observed one.
The tension is particularly relevant in the bright end of the LF, which corresponds to galaxies with UV magnitude $\muv \lesssim -20$, as most models predicted that they could form only in very massive and rare halos. This tension led to a critical revision of most theoretical models of galaxy formation, and to the proposal of several possible solutions. 

The physical processes beyond each proposed solution to the tension differ significantly.
\citet{ferrara:2023} suggested that reduced dust attenuation could explain the bright end of the LF. The absence of dust attenuation could be caused by radiation pressure due to previous intense star formation, leading to a lower column density from the same dust mass \citep{ferrara:2024, ferrara:2025c, ferrara:2025b}.
Alternatively, \citet{mason:2023} proposed that the high-$z$ LF is dominated by galaxies with the most extreme variation in the star formation rates (SFR). This requires that these extreme values are relatively common, thus requiring a stochastic SFR. Such a stochasticity may also arise from the stellar initial mass function (IMF) dependence on metallicity and redshift \citep[see][]{haslbauer:2022,trinca:2024,yung:2024,hutter:2025}, and whether stochasticity is viable or not to explain the LF tension is still under investigation \citep{pallottini:2023, sun:2023, gelli:2024, carvajal-bohorquez:2025, pallottini:2025, munoz:2026}. 
A different option is that galaxies could be intrinsically more efficient at forming stars. A high star formation efficiency (SFE) is possible if feedback from supernovae explosions does not affect star formation \citep{dekel:2023, li:2024}, or if the high density environment has a significant impact \citep{boylan-kolchin:2025, somerville:2025}, even though limits due to the impact of Lyman-$\alpha$ radiation pressure feedback have questioned the picture \citep{ferrara:2025a, manzoni:2025}.
Finally, physics beyond the standard \lcdm model may also be relevant, such as an early dark energy contribution \citep{shen:2024} or primordial black holes (PBHs, \citealt{liu:2022, matteri:2025, matteri:2025a}).
Summarizing, although several solutions have been proposed, none of them has been established as the definitive one.

Future observations can be used to probe the UV LF at even higher redshift and with greater precision, but might provide limited information to test the solutions proposed. This is due to the relatively small amount of information present in the LF, and due to different solutions giving the same LF.
Although the presence of even brighter galaxies at extreme redshift ($z\approx 25$) may exclude some solutions \citep{matteri:2025a}, an independent test would make the discrimination easier.
Indeed, investigating higher-order correlation functions of galaxy spatial distribution, which quantify the clustering properties, is an ideal candidate for such a test \citep{munoz:2023}.

The galaxy bias, a summary statistic for the clustering, has been recently measured at $z\gtrsim10$ in two independent ways.
On the one hand, \citet{dalmasso:2024, dalmasso:2026} measured the bias from JWST deep observations of the GOODS-South field, retrieving data for galaxies down to $\muv\approx -17$.
Similarly, \citet{paquereau:2025} measured the bias from the COSMOS-Web JWST survey in the redshift bin $10.5 < z < 14$ for galaxies with stellar mass $M_\star > 10^{8.85}\,\Msun$.
On the other hand, \citet{weibel:2025} measured the bias of galaxies with $\muv< -19.5$ from the cosmic variance, which is the scatter in the number of galaxies from independent lines of sight.
These data could provide an independent test for theoretical models of high-$z$ galaxy formation, eventually ruling out some of the proposed scenarios to explain the presence of bright, blue monsters at high-$z$.
Although previous works attempted such a test \citep[e.g.][]{gelli:2024, weibel:2025, chakraborty:2026}, either only a single theoretical scenario is considered, or a limited observational sample is adopted.

In this paper, we perform a systematic investigation by adopting several representative theoretical models to predict the clustering measurements, and test these results against available measurements at $z\gtrsim 10$ with the final goal of using clustering to discriminate among the various proposed scenarios. 

\section{Methods}\label{sec:methods}

The LF is the lowest order, or one-point, correlation function (CF), namely the number density of galaxies. It does not contain information on the clustering, which is quantified by higher-order CFs, based on the relative positions of galaxies. Since the number of galaxies at $z>10$ is limited, we only focus on the 2-point CF (2PCF) and its summary statistics (namely the galaxy bias), which have already been measured \citep[e.g. by][]{dalmasso:2024,dalmasso:2026,paquereau:2025,weibel:2025}.

We aim at testing different theoretical models within the standard structure formation paradigm. This means that we assume the standard $\Lambda$ cold dark matter (DM) buildup in halos, and use it as a test bench for other physical processes\footnote{In this paper, we assume a flat cosmology based on \citet{planckcollaboration:2016}. In particular, we use $\Omega_{\Lambda,0} = 0.691$, $\Omega_{m, 0} = 0.309$, $\Omega_{b, 0} = 0.0486$, $h = 0.6774$, and $\sigma_8 = 0.8159$.}.
We consider a halo catalog from a DM-only simulation to obtain a mock realization of the Universe, then we populate the halos from the catalog by adopting different theoretical models to predict both the LF and the 2PCF.
Since all the considered models are taken from scenarios meant to solve the overabundance of high-$z$ galaxies, we expect them to give similar LF, but potentially different 2PCF.

In \cref{sec:theoretical-models} we list the theoretical models considered and their main properties. In \cref{sec:dm-simulations} we present the simulation used to build the DM halo catalog. In \cref{sec:xi-computation} we explain the procedure used to compute the 2PCF from populated halos. In \cref{sec:bias-computation} we present how we computed the bias of galaxies from their 2PCF.

\subsection{Theoretical models}\label{sec:theoretical-models}

The list of solutions tested in this paper is a representative, although not exhaustive, sample of the ones present in the literature.
Each scenario gives a prescription to compute the AB absolute UV magnitude $\muv$ of a galaxy as a function of the mass of the hosting halo ($M_h$) at redshift $z$. Note that prescriptions can be stochastic, which means $\muv$ may also depend on a pseudo-random number in addition to $z$ and $M_h$.
%

\subsubsection*{Attenuation-free model}

We implement an attenuation-free model (\AFM, hereafter) by following \citet{ferrara:2023}.
The \AFM~model associates a SFR with each halo using
\begin{equation}
    {\rm SFR} = \varepsilon_\star \frac{\Omega_b}{\Omega_m} \frac{M_h}{t_{\rm ff}}\,,
\end{equation}
where $\varepsilon_\star$ is the star formation efficiency (SFE) and $t_{\rm ff}$ the free-fall time of the halo.
\AFM~assumes the SFE is regulated by supernova (SN) feedback \citep{dayal:2014}
\begin{equation}
    \varepsilon_\star = \varepsilon_0 \frac{v_c^2}{v_c^2 + f_w v_s^2}\,,
\end{equation}
where $\varepsilon_0 = 0.2$, $v_c= v_c(M_h)$ is the circular velocity of the halo, $f_w=0.12$ is the energy coupling with gas, and $v_s = \sqrt{\SI{e51}{erg} / \SI{52.89}{\Msun}} \approx \SI{975}{km/s}$ quantifies the energy released by each SN per stellar mass formed.
The free-fall time $t_{\rm ff}$ is computed as
\begin{equation}
    t_{\rm ff} = \sqrt{\frac{1}{4\pi G\rho}}\,,
\end{equation}
where $G$ is Newton's constant, and $\rho=18\pi^2 \Omega_m(z)\rho_c(z)$ is the typical average density of halo formed at $z$. Note that no stochasticity is included in the SFR computation.
Finally, the SFR of each galaxy is converted into a UV luminosity through
\begin{equation}\label{eq:AFM-luminosity}
    L\ped{\rm UV} = \mathcal{K} \ {\rm SFR}\,,
\end{equation}
where $\mathcal{K} = \SI{0.587e10}{\Lsun / (\Msun / yr)}$.

The key part in \AFM is the absence of dust attenuation at high-$z$, which could decrease the measured luminosity by orders of magnitude. This affects the LF by making each galaxy brighter, while letting the observed luminosity grow with the halo mass.

\subsubsection*{Feedback-free burst}

We implement the feedback-free burst model (\FFB, hereafter) by following \citet{li:2024}.
Because of the high ISM density, the time scale for star formation may become shorter than the SN feedback, making it less effective; in practice, \FFB~allows galaxies to have SF episodes that can be considered feedback-free (FF), thus effectively increasing the star formation efficiency.
These episodes enhance the stellar mass with respect to typical post-starburst scenarios, leading to more massive and brighter galaxies.

\citet{li:2024} based their SF prescription\footnote{The implementation is publicly available via \href{https://github.com/syrte/ffb\_predict}{this} repository.} on UniverseMachine \citep{behroozi:2019}. They define ${\rm SFR}_{um,sf}(M_h, z)$ (${\rm SFR}_{um,q}(M_h, z)$) as the SFR of star-forming (quenched) galaxies, based on Appendix H in \citet{behroozi:2019}.
On top of this, a fraction $f_{\rm ffb}$ of SF galaxies are considered to be in a FF burst, with
\begin{equation}
    \langle {\rm SFR}_{\rm ffb}\rangle = \varepsilon_{max} \frac{\Omega_b}{\Omega_m} s M_h \left(\frac{M_h}{\SI{e12}{\Msun}}\right)^{0.14} (1+z)^{2.5}\ ,
\end{equation}
where $s=\SI{0.03}{Gyr^{-1}}$. We take $\varepsilon_{max}=0.15$ to match the LF at $z \approx 11$, as suggested by \FFB, and verified in this work (see \cref{fig:luminosity-function-check}).

\citet{li:2024} takes the fraction $f_{\rm ffb}$ as a smoothed version of the one by \citet{dekel:2023}, who give a threshold on the halo mass for the FF burst
\begin{equation}
    M_{h, \rm ffb}= 10^{10.8}\,\Msun \,\left(\frac{1+z}{10}\right)^{-6.2}\ .
\end{equation}
In particular,
\begin{equation}
    f_{\rm ffb} = \mathcal{S}\left(\frac{\log(M_h/M_{h, \rm ffb}(z))}{\Delta {\log M}} \right)\ ,
\end{equation}
where $\mathcal{S}(x) = (1+e^{-x})^{-1}$ and $\Delta {\log M} = 0.15$.

For each galaxy, the SFR is integrated to retrieve the stellar mass. Since \citet{behroozi:2019} provides self-consistently $M_{\star,um} = M_{\star,um}(M_h,z)$ relations, the following integration is performed 
\begin{equation}
    M_\star = M_{\star,um} + \int_0^{t(z)}\langle {\rm SFR}_{um+ \rm ffb} - {\rm SFR}_{um}\rangle \dd t'\ ,
\end{equation}
where the SFR in the integral depends on $M_h(t')$, which is assumed to follow a typical halo assembly history \citep{dekel:2013}. On top of the average value of $M_\star(M_h, z)$, \FFB considers a lognormal scattering with
\begin{equation}
    \sigma_\star = 0.1 + 0.3 \mathcal{S}\left(\frac{13.9 - 0.3 z - \log M_h/\Msun}{0.2}\right)\,.
\end{equation}
Then, $\muv$ is stochastically computed from the stellar mass \citep{yung:2024}, assuming a Gaussian distribution with mean
\begin{equation}
    \langle\muv\rangle = -2.3 \log \frac{M_\star}{10^9 \,\Msun} - 20.5
\end{equation}
and standard deviation $\sigma_{\muv} = 0.3$. Finally, we apply a dust correction following the ``disc'' prescription by \citet{li:2024}.

The key part in \FFB~is that a fraction of galaxies in massive halos ($\gtrsim M_{h,\rm ffb}$) is significantly brighter than the others, thus boosting the bright-end of the LF. Smaller halos host significantly fainter galaxies and contribute negligibly to the LF.

\subsubsection*{Bursty star formation}

We implement a bursty SF model (\BSF, hereafter) by following \citet{gelli:2024}.
\BSF computes the typical UV magnitude (${\rm M_{\rm UV,c}}$) of galaxies hosted in halos of mass $M_h$, and applies a Gaussian scattering on top of this value.
This means that the magnitude of each galaxy is sampled from a Gaussian with mean ${\rm M_{\rm UV,c}}$ and standard deviation
\begin{equation}
    \sigma_{\rm UV} = -0.34 \log \left(\frac{M_h}{M_\odot}\right) + 4.5\, .
\end{equation}

In \BSF, ${\rm M_{\rm UV,c}}$ is computed from SF prescription by \citet{mason:2015}, which assumes a redshift-independent SFE
\begin{equation}\label{eq_BSF}
    \varepsilon \equiv \frac{\rm SFR}{M_h}\Delta t_h\,,
\end{equation}
with $\Delta t_h$ being the halo assembly time, that is defined as the time passed since the halo progenitor had mass $M_h/2$. Such a time scale is computed following \citet{lacey:1993}
\begin{equation}
    \Delta t_h = t_0 \left(1 + \frac{1.2}{\delta_c} \sqrt{\sigma\left(\frac{M_h}{2}\right)^2 - \sigma(M_h )^2} \right)^{-1.5}\,,
\end{equation}
where the redshift dependence is implicit: $t_0=t_0(z)$ is the age of the Universe at $z$, $\delta_c \approx 1.686$ the critical linear overdensity, and $\sigma = \sigma(M, z)$ is the mass variance on a scale that on average contains a mass $M$ at redshift $z$.
\citet{mason:2015} calibrated $\varepsilon$ with observations at $z\approx 5$.

To mimic this approach, we take the efficiency $\varepsilon= \varepsilon(M_h)$ (eq. \ref{eq_BSF}) to compute the expected SFR of the galaxy in each halo. We consider the SFR as constant over the halo assembly time, and use Starburst99 \citep{leitherer:1999} data to convert it to ${\rm M_{\rm UV,c}}$.
Due to the application to high-$z$, we choose the lowest metallicity available ($Z = 10^{-3}$) and apply no dust attenuation, as per \citet{gelli:2024}.

The key part in \BSF~is that the bright UV galaxies in the LF are reproduced via the stochasticity of the SFR, which makes a fraction of otherwise relatively dim galaxies temporarily shine very brightly. This means that bright galaxies are typically hosted in smaller halos with respect to other theoretical models.

\subsubsection*{Primordial black holes}

We implement a PBH model (\PBH, hereafter) by following \citet{matteri:2025}.
In \PBH, in each halo, the UV luminosity is given by a sum of stellar and PBH contributions, supposing each PBH can shine as an active galactic nucleus (AGN). The stellar contribution is computed from the relation
\begin{equation}
    \muv = p_1 + \left[p_2\left(\frac{M_h}{\SI{e11}{\Msun}}\right)^{p_3} + p_4\right]\log\left(\frac{M_h}{\SI{e11}{\Msun}}\right) + p_5 z^{p_6}
\end{equation}
with the $p_i$ parameters being fitted using $z<10$ UV LF data \citep{matteri:2025}
\begin{equation}
    \begin{pmatrix}
        p_1 \\p_2 \\p_3 \\p_4 \\p_5 \\p_6
    \end{pmatrix}
    =
    \begin{pmatrix}
        -13.58 \\-8.80 \\-0.0326 \\5.68 \\-3.44 \\0.299
    \end{pmatrix}
    \ .
\end{equation}
For the PBH contribution, each halo above a threshold mass $M\ped{min}=10^{7.5}\, \Msun$ has a probability to host a PBH. Such a probability is computed at each redshift by fixing the comoving number density of PBHs.
Since the probability is assumed to be independent of the halo mass, most PBHs are hosted in halos with masses slightly above $M\ped{min}$ due to the rapid decrease of the HMF with $M_h$.
The model considers each PBH as an AGN shining at the Eddington luminosity, and uses the bolometric correction from \citet{shen:2020} to compute the UV luminosity.

Each PBH mass is extracted from a lognormal distribution
\begin{equation}\label{eq:pbh-lognormal}
    \deriv{n(>M)}{\log M} \propto \frac{1}{\sqrt{2\pi}\spbh} \exp\left[-\frac12\left(\frac{\log M/\Mpbh}{\spbh}\right)^2\right]\,,
\end{equation}
with $\Mpbh = \SI{2e5}{\Msun}$ and $\spbh = 0.58$, compatible with \citeauthor{matteri:2025}'s results.
The normalization in \cref{eq:pbh-lognormal} is fixed so that the global density of the PBHs in the Universe is a fraction $\fpbh = 3.8\times 10^{-8}$ of the DM density.

The key part in the \PBH model is that the boost in luminosity comes from galaxies that are hosting PBHs. This means that a significant fraction of the bright galaxies is hosted in halos with $M_h\approx M_{min}$, where most PBHs reside.

\subsection{Dark matter simulations}\label{sec:dm-simulations}

\begin{figure}
    \centering
    \includegraphics[width=\linewidth]{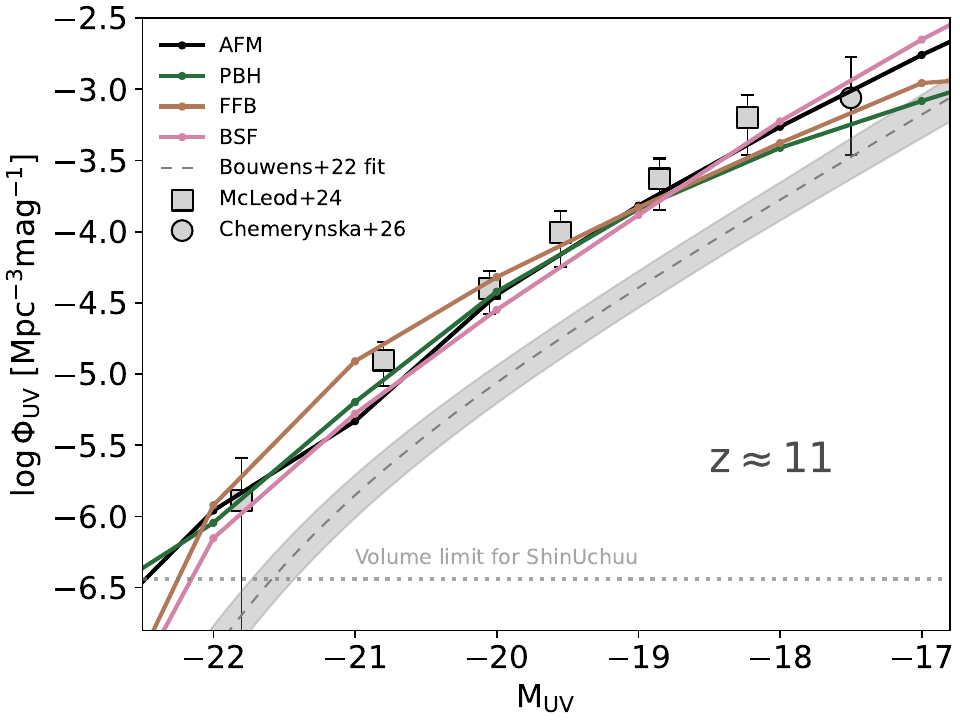}
    \caption{
        Luminosity functions (LF) at $z\approx 11$ predicted by the attenuation-free (\AFM), feedback-free burst (\FFB), bursty star formation (\BSF), and primordial black hole (\PBH) models (see \cref{sec:theoretical-models} for details).
        Each model adopts the dark matter halo population from the Shin-Uchuu simulation \citet{ishiyama:2021} at $z=10.64$, with the gray dotted line marking its volume limit (one galaxy per magnitude).
        The LF is computed on intervals centered at integer $\muv$ and with a $1 \, \rm mag$ full width.
        Filled symbols represent LF measurements at redshift $z\approx 11$ by \citet{mcleod:2024} and \citet{chemerynska:2026}, while the gray dashed line and shaded region represent a pre-JWST prediction extrapolated from \citet{bouwens:2022} fit.
        \label{fig:luminosity-function-check}
    }
\end{figure}

\begin{figure}
    \centering
    \includegraphics[width=\linewidth]{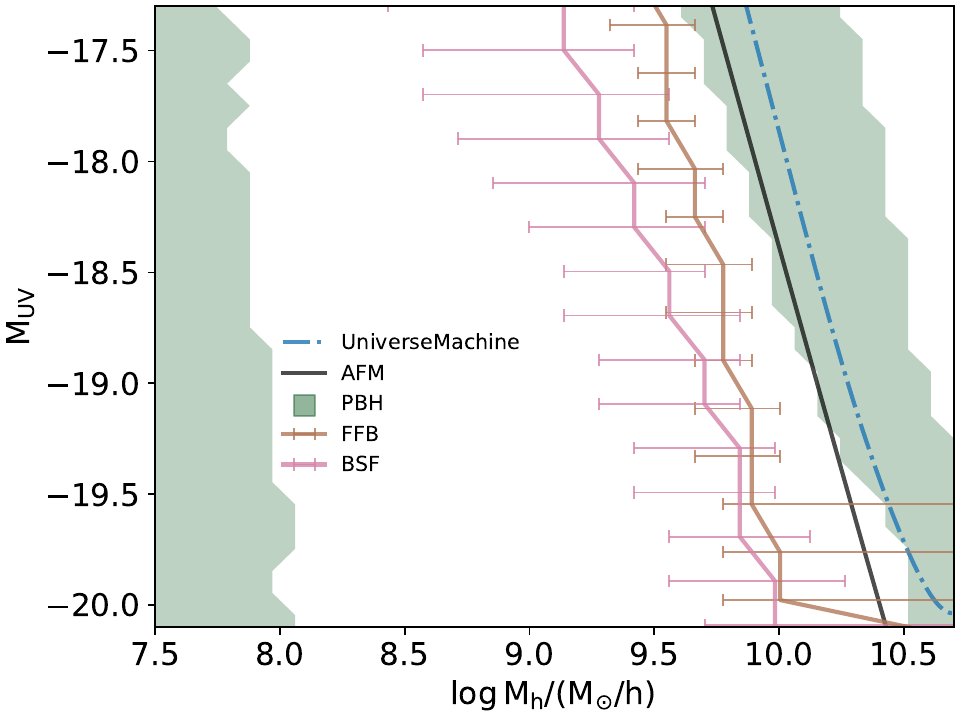}
    \caption{
        Distribution of galaxies in the host halo mass ($M_h$) vs UV magnitude ($\muv$) plane.
        Each color represents where a specific model predicts galaxies lie at $z\approx11$.
        \AFM is reported as a line, since no scatter is included in the model.
        \FFB and \BSF are plotted as lines representing the median and scatter, the latter being given by the 10th and 90th percentiles, of the halo mass associated to a given $\muv$.
        \PBH distribution is bimodal, and the shaded areas represent the location where $80\%$ of the galaxies lie.
        For reference, we additionally plot a $z\approx 11$ extrapolation based on UniverseMachine \citep{behroozi:2019}.
        \label{fig:hm-muv}
    }
\end{figure}

To sample the luminosity function appropriately, we need a simulation with a large box size and a high DM resolution to probe the bright and the faint end, respectively.
Indeed, bright galaxies are rare - thus finding them requires a large volume - and faint galaxies typically reside in small halos, which need a high resolution to be tracked.

We adopt the Shin-Uchuu simulation \citep{ishiyama:2021}, that has a good trade-off between box size ($L_{box} = \SI{140}{cMpc/h}$) and a mass resolution ($M_{\rm DM} = \SI{8.97e5}{\Msun /h}$).
From the observed LF provided by \citet{mcleod:2024}, such a volume is expected to contain roughly one galaxy with $-22.5 \leq\muv\leq -21.5$ at $z\approx 11$.
With $M_{\rm DM} = \SI{8.97e5}{\Msun /h}$, the simulation can track with $\approx 100$ particles halos with mass $M_h\approx 10^8\,\Msun$, i.e. about the mass expected for atomic-cooling halos.
Given a halo catalog from Shin-Uchuu at a specific redshift, we compute the $M_{\rm UV}$ predicted by each model (\cref{sec:theoretical-models}), and build the corresponding LF.

As a first check, in \cref{fig:luminosity-function-check} we plot the resulting LFs at $z=10.64$, the first after $z=11$ from Shin-Uchuu.
As expected, all the models predict LFs that are consistently above the pre-JWST prediction (gray dashed line and shaded region, \citealt{bouwens:2022}), and consistent with $z\approx 11$ JWST measurements (gray symbols, \citealt{mcleod:2024,chemerynska:2026}).
Although the LFs from all theoretical models are almost indistinguishable, the distributions of galaxies are very different when plotted in the $\log M_h$-$\muv$ plane, as can be appreciated from \cref{fig:hm-muv}.
Models in which $\muv$ is deterministic (e.g. \AFM) are reported as a line, whereas for models with stochastic prescriptions (e.g. \BSF and \FFB) we report the areas including $80\%$ of galaxies of a given $\muv$.
Note that the shape of \PBH's distribution is significantly different because of the bimodal nature of galaxies considered: massive galaxies without PBHs or small galaxies hosting a PBH AGN.
For all considered models, most of the galaxies are predicted to lie below the low-$z$ extrapolation from pre-JWST data, here represented by the blue line \citep[UniverseMachine,][]{behroozi:2019}.
Indeed, to match $z>10$ LF from JWST, models have to place a significant amount of bright galaxies in smaller and thus more abundant halos when compared to pre-JWST models.

Qualitatively, we can understand several properties of different models from \cref{fig:hm-muv}.
With respect to the others, \AFM predicts the brightest galaxies are hosted in the most massive halos (e.g. $M_h(\muv= -19) \approx 10^{10.2}\,\Msun/h$), thus we expect the scenario to give the highest clustering prediction, as shown later in \cref{sec:results}.
\FFB and \BSF distributions are both stochastic, but have different medians and scatter. \BSF features a lower median $M_h$ at fixed $\muv$ and a larger scatter, e.g. galaxies with $\muv\approx -19$ are placed in halos with $\log \frac{M_h}{\Msun/h} = 9.7^{+0.1}_{-0.4}$ by \BSF and $\log \frac{M_h}{\Msun/h} = 9.8^{+0.1}_{-0.1}$ by \FFB. Since these models retain the same LF, the lower median is effectively compensated by the higher scatter.
Given the lower typical halo mass associated with a given $\muv$, galaxies from \BSF and \FFB are expected to be less clustered than those from the \AFM.
Finally, the \PBH distribution is bimodal and predicts most bright galaxies (roughly 3 out of 4, \citealt{matteri:2025}) are PBHs shining in low mass halos ($M_h \lesssim 10^8\,\Msun$), thus placing them in a more uniform and less clustered way.

\subsection{$\xi(r)$ computation}\label{sec:xi-computation}

The two-point CF (2PCF) is defined as \citep[see e.g. ][]{peebles:1980}
\begin{equation}
    \xi(r) = \frac{p_D(r)}{p_R(r)} - 1\,,
\end{equation}
where $p_D(r)$ is the probability of finding a galaxy at a distance $r$ from another one, and $p_R(r)$ is the same probability in the case of uniformly distributed galaxies. Typically, the 2PCF is decreasing with $r$ on scales $\approx0.1-10 \, \rm cMpc / h$, being well-fitted by a power-law with negative index \citep{peebles:1980}. In a \lcdm model, this is caused by large-scale fluctuations in the DM field that cluster massive halos. Since these halos foster galaxy formation, galaxies tend to be observationally clustered.
If the brightest galaxies were hosted in the most massive halos, then brighter galaxies would be more clustered. This hypothesis may not be true, for example, due to the presence of massive passive galaxies at $z \lesssim 4$ \citep{wechsler:2018}.
Nonetheless, measurements show that clustering positively correlates with brightness \citep{zehavi:2011}.

Each theoretical model (\cref{sec:theoretical-models}) yields a prescription to populate halos with galaxies (\cref{fig:hm-muv}). By assuming that galaxies are placed at the center of DM halos in the Shin-Uchuu simulation, we can build a catalog of galaxy positions and $\muv$ magnitudes.

We filter such a catalog by removing galaxies below a $\muv$ threshold. Different $\muv$ thresholds are tested in order to compare with different luminosity-limited observations.
Then, we compute the 2PCF $\xi(r)$ from such a flux-limited sample. This procedure is required to mimic the flux-limited data from photometric observations, and to account for the incompleteness of faint sources \citep{dalmasso:2024}.
We remark that this prevents us from comparing directly our results with mass-limited observations \citep[e.g.][]{paquereau:2025}. 

Following \citet{landy:1993}, we take the subsample of galaxies we are interested in (hereafter $D$) and perform pair counts to estimate the 2PCF $\xi(r)$.
Firstly, we build a supplementary random catalog (hereafter $R$) assuming a uniform distribution. This catalog represents a realization of a box with the same geometry, but with a 25 times higher number density\footnote{We chose this number following \citet{mohammad:2022}. In addition, to avoid problems when very few galaxies are present in the $D$ catalog, we impose that the $R$ catalog contain at least $10^6$ points.}.
Secondly, we count the number of pairs that are separated by a distance $r$. These pairs may reside either in the same catalog (both in $D$ or both in $R$) or in two different catalogs (one point in $R$ and one in $D$), and are named accordingly, i.e. $DD$, $RR$, and $DR$.
Thirdly, the pair counts are normalized by the total number of possible couples\footnote{The normalization is $N(N-1)$ for a pair in a catalog of size $N$, while it is $N_1 N_2$ for a cross pair from two distinct catalogs of sizes $N_1$ and $N_2$.}.

Then, the 2PCF can be computed with the \citeauthor{landy:1993} (or LS) estimator:
\begin{equation}\label{eq:ls-xi-estimator}
    \xi(r) = \frac{DD(r) - 2DR(r) + RR(r)}{RR(r)}\,.
\end{equation}
Albeit having a higher computational cost, LS typically performs better than other estimators \citep{kerscher:2000}.
In using \cref{eq:ls-xi-estimator}, we bin the distances in $4$ logarithmically spaced bins ranging $0.1 \leq r_i/{{\rm cMpc}\,h^{-1}} \leq 10$.
Each pair count is computed for distances in the bins, retrieving $4$ real numbers for each couple of selected catalogs.

To estimate the uncertainty for $\xi(r)$, we follow the jackknife (JK) procedure described by \citet{mohammad:2022}.
Given a cubic volume of side $L$, we divide it into $N_{JK}^3=27$ sub-cubes of side $L/N_{JK}$ and recompute $\xi(r)$ while removing each time a different sub-cube.
As proposed by \citet{mohammad:2022}, we use a ``match'' weighting scheme in the pair counting to correct the Poisson noise on $\xi$ with the JK. This means that we assign a weight $\alpha$, which depends only on $N_{JK}$, to a pair constituted by a galaxy in the removed sub-cube and one out of it, instead of simply neglecting the pair.
Such a scheme is effective when removing even a few galaxies with the JK makes the $\xi$ change significantly, that is, when the galaxy density is low. We remark that it may perform poorly at high galaxy densities, when the Poisson noise is subdominant \citep{trusov:2024}.

The JK procedure gives $N_{JK}^3$ different estimates of $\xi(r)$, which we label $\xi_j(r_i)$, with $j$ being JK index from 1 to $N_{JK}^3$, and $r_i$ a selected radial bin. We take
\begin{equation}
    \bar\xi(r_i) = \frac{1}{N_{JK}^3}\sum_{j=1}^{N_{JK}^3} \xi_j(r_i)
\end{equation}
as the mean value for $\xi(r_i)$, and compute the covariance\footnote{A posteriori, if only the diagonal terms of $C_{ab}$ are kept, the uncertainties on the results reported in \cref{sec:results} would be underestimated by a factor $\approx2$.} of $\xi$ as
\begin{equation}
    C_{ab} = \frac{N_{JK}^3-1}{N_{JK}^3}\sum_{j=1}^{N_{JK}^3}(\xi_j(r_a) - \bar\xi(r_a))(\xi_j(r_b) - \bar\xi(r_b))\ .
\end{equation}

Since on the scales of interest here ($0.1-10 \,\rm{cMpc}\,h^{-1}$), $\xi(r)$ is expected to behave as a power-law \citep{peebles:1980}, we adopt as our fitting function
\begin{equation}\label{eq:xi-power-law}
    \xi = \left(\frac{r_0}{r}\right)^{\gamma}
\end{equation}
where $\gamma$ and $r_0$ are parameters. We perform the fit considering the full covariance matrix retrieved from the JK procedure.

\begin{figure}
    \centering
    \includegraphics[width=\linewidth]{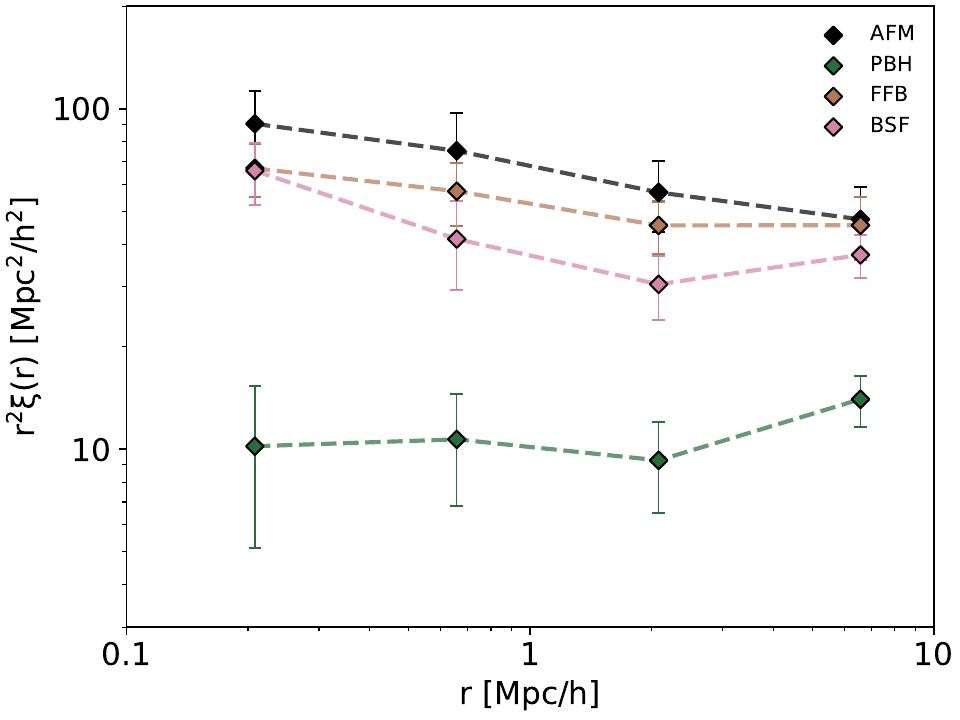}
    \caption{
        The 2-point correlation function ($\xi$, 2PCF, \cref{eq:ls-xi-estimator}) of $z\approx 11$ galaxies brighter than $\muv=-19$ for all models (see \cref{sec:theoretical-models}).
        For each model, $\xi$ is fitted with a power-law to compute the corresponding bias (\cref{sec:bias-computation}).
        For visualization sake, $\xi$ is plotted multiplied by $r^2$, so that a rising (descending) line means that the negative power-law index ($\gamma$, \cref{eq:xi-power-law}) is smaller (larger) than 2.
        \label{fig:xi-example}
    }
\end{figure}

To visualize the differences between models, in \cref{fig:xi-example} we report the values of $r^2\xi(r)$ for galaxies brighter than $\muv=-19$ from our approach.
Both the normalization and the slope are tracers of the clustering: the higher (or the steeper) the curve in the plot, the more clustered the galaxies are.
On the one hand, the slope ($\gamma$) is very similar in all cases, e.g. from \cref{fig:xi-example} we get $\gamma\approx 2.15$ from \AFM, \BSF, and \FFB, while $\gamma\approx 1.81$ from \PBH.
On the other hand, the normalization varies much more, allowing a direct comparison of the models: the \AFM predicts the highest clustering, followed in order by \FFB, \BSF, and \PBH, as expected from the $\muv-M_h$ relation (\cref{sec:dm-simulations}).

\subsection{Bias computation}\label{sec:bias-computation}

Since galaxies are clustered, considering them to infer the underlying DM distribution nets a biased result \citep{kaiser:1984, desjacques:2018a}. The 2PCF of halos and galaxies is typically compressed into the bias measure
\begin{equation}\label{eq:bias_definition}
    b = \sqrt{\frac{\xi_g(r)}{\xi\ped{\rm DM}(r)}}\ ,
\end{equation}
where $\xi_g$ ($\xi\ped{\rm DM}$) is the 2PCF of galaxies (Dark Matter).
Using the definition in \cref{eq:bias_definition}, the bias is independent of $r$ on scales large enough to neglect the non-linear contributions to the DM distribution.

Equivalently to \cref{eq:bias_definition}, the bias can be written in terms of the power spectrum, since the 2PCF is the Fourier transform of such a quantity. Then, given the linearity of the Fourier transform, the equivalent of \cref{eq:bias_definition} in $k$-space is
\begin{equation}\label{eq:bias_PS}
    b = \sqrt{\frac{P_g(k)}{P\ped{\rm DM}(k)}}\,,
\end{equation}
where $k$ is a sufficiently small wavenumber ($k\lesssim 0.2 \, \si{h/Mpc}$, \citealt{desjacques:2018a}), and $P_g$ ($P\ped{\rm DM}$) is the linear power spectrum of galaxies (dark matter).
Since the bias does not depend on $k$, the only difference in the two power spectra is the normalization. This is typically defined as the linear fluctuation on a scale of $r=\SI{8}{h^{-1}cMpc}$, and for the dark matter this is the $\sigma_{8}$ cosmological parameter. For galaxies, we can compute \citep{peebles:1980}
\begin{equation}
    \sigma_{8,g}^2 = \frac{72 (r_0 / \SI{8}{h^{-1} Mpc})^\gamma}{(3-\gamma)(4-\gamma)(6-\gamma)2^\gamma}
\end{equation}
from the parameters obtained from \cref{eq:xi-power-law}.
Finally, remembering that $P_i \propto\sigma_i^2$, the bias can be computed as
\begin{equation}\label{bias_xi}
    b = \frac{\sigma_{8,g}}{\sigma_{8}}\ ,
\end{equation}
where $\sigma_{8} = \sigma_{8}(z)$ is the cosmological parameter accounting for the linear growth factor of fluctuations.
The uncertainty on $b$ is computed by propagating the error for $\gamma$ and $r_0$.


\section{Results}\label{sec:results}

\begin{figure*}
    \centering
    \includegraphics[width=0.9\linewidth]{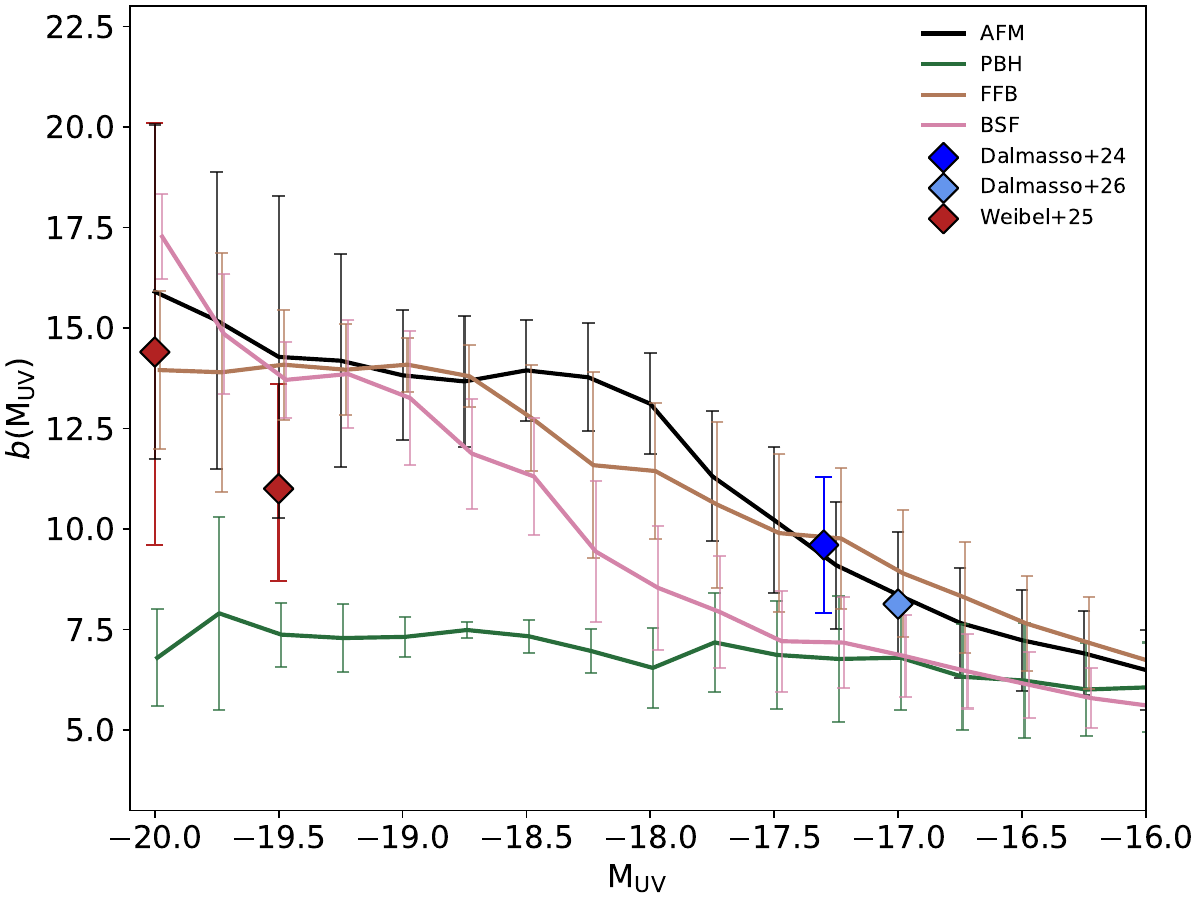}
    \caption{
        Bias estimate for all the adopted models (see \cref{sec:theoretical-models}) as a function of the UV magnitude limit.
        The bias is computed by fitting the 2-point Correlation Function $\xi$ (see \cref{sec:xi-computation} and \cref{sec:bias-computation}), also propagating the uncertainties from the fit.
        We also report observational data from \citet{dalmasso:2024, dalmasso:2026} and \citet{weibel:2025}.
        \label{fig:bias_all}
    }    
\end{figure*}

\begin{table*}[ht]
    \centering
    \caption{Bias observations and predictions}
    \begin{tabular}{c|c|c|c|c|c}
        \toprule
        Source & $b(\muv < -20)$ & $b(\muv < -19.5)$ & $b(\muv \lesssim -17.3)$ & $\chi^2$ (3 dof) & p-value\\
        \midrule
        \citet{dalmasso:2024} & / & / & $9.6^{+1.7}_{-1.7}$ & / & / \\
        \citet{dalmasso:2026} & / & / & $8.13^{+0.04}_{-0.02}$ & / & / \\
        \citet{weibel:2025}   & $14.4^{+5.7}_{-4.8}$ & $11.0^{+2.6}_{-2.3}$ & / & / & / \\
        \AFM & $16.0^{+4.0}_{-4.0}$ & $14.0^{+4.0}_{-4.0}$ & $9.1^{+1.6}_{-1.6}$ & $0.49$ & 0.08\\
        \BSF & $17.3^{+1.1}_{-1.1}$ & $13.7^{+0.9}_{-0.9}$ & $7.2^{+1.1}_{-1.1}$ & $2.62$ & 0.45\\
        \FFB & $14.0^{+2.0}_{-2.0}$ & $14.1^{+1.4}_{-1.4}$ & $9.8^{+1.7}_{-1.7}$ & $1.14$ & 0.23\\
        \PBH & $6.8^{+1.2}_{-1.2}$  & $7.4^{+0.8}_{-0.8}$  & $6.8^{+1.6}_{-1.6}$ & $5.98$ & 0.11\\
        \bottomrule
    \end{tabular}
    \label{tab:bias-results}
\end{table*}

To mimic flux-limited observational surveys, we compute the bias of galaxies at $z\approx 11$ applying different $\muv$ thresholds\footnote{The bias is computed up to $\muv=-20$ since the estimation fails for brighter objects, due to the limited number ($\approx 150$) of simulated halos.} (every $0.25\,\rm mag$ from $-20$ to $-16$).

\cref{fig:bias_all} and \cref{tab:bias-results} show our results, along with measurements by \citet{dalmasso:2024, dalmasso:2026} and \citet{weibel:2025}\footnote{We remark that predictions in the fourth column of \cref{tab:bias-results} are computed for galaxies with $\muv < -17.25$ because of our binning scheme, whereas the \citet{dalmasso:2026} measurement is for $\muv<-17$.}.
For faint galaxies ($\muv \approx -16$), all the models produce roughly the same bias ($b\approx 7$), making them virtually indistinguishable.
As luminosity increases, while the \PBH model predicts a constant bias, all the other models predict that it grows, with relatively small differences among them.
As we expected from \cref{sec:dm-simulations}, the general trend sees the \AFM model predicting a high bias, followed by \FFB and \BSF, while \PBH gives the overall smallest prediction.

The bias from observations increases slowly with luminosity, remaining potentially compatible with a constant. This growth seems slower than predicted by \BSF, \FFB, and \AFM, while it is faster than predicted by \PBH.
However, the uncertainties on both predictions and measurements are relatively large, meaning that all models are consistent within 1.5-sigma with all observations. For this reason, no clear tension appears, and no model can be firmly ruled out based on the analysis with current data.

We remark that the low luminosity constraint by \citet{dalmasso:2026} adopted a Halo Occupation Distribution (HOD) method to recover the bias, instead of the power-law fit we performed. The resulting $\xi(r)$ is within the bulk part of the scenario analyzed in the present paper, but the uncertainty is significantly smaller.
Even when based on the same $\xi(r)$, these approaches can give incompatible results (see section 4.2 of \citealt{dalmasso:2026}). As the cause appears to reside in the different methods used, we decided not to use this measurement in our statistical analysis.
We also do not compare with measurements at luminosities higher than $\muv = -20$, like the brightest measurement by \citet{weibel:2025} and the $z\approx12$ one by \citet{paquereau:2025}\footnote{Although the measurement by \citet{paquereau:2025} is mass-limited, an abundance matching associates a flux limit of roughly $\muv<-20.5$ to their $z\approx12$ point (see their Table 1).}, as we have no predictions in that regime.

We compute the $\chi^2$ between each model and observations (fifth column of \cref{tab:bias-results}), measured as the sum of $(b_{pred} - b_{obs})^2/(\sigma^2_{pred} + \sigma^2_{obs})$ over the 3 samples, and assume all predictions are statistically independent.
From the combination of observations at different $\muv$, we do not get a significant tension with predictions, as all p-values are above $0.05$.
This means that we cannot rule out any of the tested models.
Although this result could be improved by taking advantage of additional independent simulation snapshots to reduce the uncertainty in predictions, observations also put limits on the room for improvement.
For example, if we arbitrarily set the uncertainty on predictions to 0, only a small tension would arise with the p-value of \PBH reaching $0.05$, while the other p-value would remain substantially unchanged.
This means that both a theoretical and observational effort is required to use the bias as a probe to statistically discriminate among various models.

The most significant difference between predictions is the $\muv$ at which each curve reaches the high-luminosity bias value ($b\approx14$). This means additional observations at $\muv\lesssim -18$ could be crucial to tighten the constraints on the models, potentially fixing the shape of $b(\muv)$.
Also at higher luminosities ($\muv<-20$) models could give different predictions, as the high biases give stricter constraints on the associated halo mass \citep[see e.g.][]{tinker:2010}.

\section{Conclusions}

In this paper, we exploited current clustering measurements to systematically test the scenarios that aim at solving the overabundance problem of bright blue super-early ($z\geq 10$) galaxies.
We have chosen 4 representative theoretical prescriptions that reproduce the observed $z\approx 11$ UV luminosity function (LF, \cref{fig:luminosity-function-check}), namely the Attenuation Free Model (\AFM), the Feedback Free Burst model (\FFB), a Bursty Star Formation prescription (\BSF), and a Primordial Black Holes model (\PBH).
Adopting the Shin-Uchuu simulation \citep{ishiyama:2021}, we have populated halos at $z\approx 11$ to predict the galaxies' emission in different scenarios. From the catalog of galaxies, we have computed the 2-point Correlation Function (2PCF) $\xi(r)$, fitted it with a power-law, and used the results to predict the galaxy bias from each model. Such predictions of the bias have been compared with current observations by \citet{dalmasso:2024, dalmasso:2026} and \citet{weibel:2025}.

We find that the main discriminating power arises from galaxies brighter than $\muv=-18$ (\cref{fig:bias_all}), for which different models assign galaxies to substantially different halo masses (\cref{fig:hm-muv}), and therefore predict distinct clustering amplitudes. At fainter magnitudes, instead, the predicted biases converge, making the models effectively indistinguishable with current data.
Despite these differences, current observational and theoretical uncertainties remain too large to confirm which scenario best describes the data.
In particular, no model presents a significant tension (p-value $<0.05$) when compared with observations (\cref{tab:bias-results}).

On the theoretical side, further progress will require reducing the uncertainties in model predictions.
A larger simulation volume would provide more massive halos, thus typically improving the predictions on the bias of bright galaxies. Nonetheless, this would mean sacrificing mass resolution, required to trace the small mass halos that typically host faint galaxies.
Among the cosmological simulations with publicly available data, Shin-Uchuu provides one of the best compromises for our purposes.
Future studies should consider using additional independent snapshots from other simulations to provide a more numerous statistical sample and reduce uncertainties.

On the observational side, significant improvements are expected in the near future.
For example, \citet{weibel:2025} based their estimate on 34 independent JWST lines of sight and argued that adding 50 more could reduce their current $2\sigma$ uncertainties to approximately the present $1\sigma$ level.
If achieved, such a gain in precision could substantially strengthen the tension with some of the models considered here.
Likewise, larger contiguous surveys would improve clustering measurements based on the power-law fit to the 2PCF, potentially probing the knee of the $b(\muv)$ function around $\muv=-18$, and taking advantage of the differences in model predictions. 
As an example, an area of $\SI{233}{arcmin^2}$ is expected to yield roughly 200 galaxies with $\muv\lesssim-18$ at $z=10.5$-$11.5$, four times the sample used by \citet{dalmasso:2024} for their $z>10$ estimate.
Assuming the error scales as the square root of the number of galaxies, such a survey should give an uncertainty close to $\pm 1$ on the bias and enhance the tension between models, given their predictions span the range between $b\approx 7$ (\PBH) and $b\approx 14$ (\AFM).

In this perspective, both \textit{Euclid} and \textit{Roman} could play an important complementary role by probing larger portions of the sky.
\textit{Euclid} could detect galaxies as bright as $\muv\approx -21.5$ in fields such as the Euclid Deep Fields \citep{euclidcollaboration:2025a, euclidcollaboration:2025}. Although these luminosities lie beyond the range explored in this paper, the clustering of such rare and luminous systems could provide one of the clearest observational tests of the different physical scenarios proposed to explain the excess of bright galaxies at very high redshift.
\textit{Roman} could also efficiently probe the 2PCF at $z \gtrsim 10$ by detecting a large number of $\muv <-17$ galaxies in the Ultra Deep Field \citep{drakos:2022}.

In summary, current clustering measurements do not yet allow us to distinguish among the proposed solutions to the JWST high-$z$ luminosity function tension.
However, our analysis shows that galaxy bias, particularly at the bright end, has the potential to become a powerful and independent discriminator of early galaxy formation models.
Fully exploiting this potential will require both more precise observations and improved theoretical predictions.

\begin{acknowledgements}
AF acknowledges support from the ERC Advanced Grant INTERSTELLAR H2020/740120.
We gratefully acknowledge the computational resources of the Center for High Performance Computing (CHPC) at SNS.
We acknowledge usage of the \code{Python} programming language, \code{Astropy} \citep{astropycollaboration:2013, astropycollaboration:2018, astropycollaboration:2022}, \code{Corrfunc} \citep{sinha:2019, sinha:2020}, \code{hmf} \citep{murray:2013}, \code{Matplotlib} \citep{hunter:2007}, \code{Multiprocess} \citep{mckerns:2010, mckerns:2012}, \code{Numba} \citep{lam:2015}, \code{NumPy} \citep{harris:2020}, and \code{SciPy} \citep{virtanen:2020}.
\end{acknowledgements}

\bibliographystyle{aa_url}
\bibliography{main}

\end{document}